\documentclass[useAMS,usenatbib]{mn2e}

\usepackage{epsfig, natbib, graphicx, color}

\input epsf 

\usepackage{color}

\usepackage[english]{babel}
\usepackage{graphicx}
\usepackage{xcolor}
\usepackage{amsmath}
\usepackage[overload]{empheq}
\usepackage{array}
\usepackage{indentfirst}
\usepackage{perpage} %the perpage package
\MakePerPage{footnote} %the perpage package command
\newcommand\prd[3]{ {Phys. Rev. D}} 
\newcommand\PhysRevD[3]{ {Phys. Rev. D}} 
\newcommand\PhysRevLet[3]{ {Phys. Rev. Letters} }

\newcommand{\eg}{\hbox{{\it e.g.},}}

\newcommand{\lesssim}{\let\la}
\newcommand{\gtrsim}{\let\ga}

\newcommand*{\pablo}[1]{\textcolor{red}{\textsf{Pablo: #1}}}

\newcommand{\boldalpha}{\boldsymbol{\alpha}}
\newcommand{\boldpi}{\boldsymbol{\pi}}

\newcommand{\LCDM}{$\Lambda$CDM}

\newcommand{\msol}{M_\odot}

\newcommand{\none}{n_1}
\newcommand{\ntwo}{n_2}
\newcommand{\nthree}{n_3}

\newcommand{\betaone}{\beta_1}
\newcommand{\betatwo}{\beta_2}
\newcommand{\betathree}{\beta_3}

\newcommand{\rab}{r_{ab}}

\newcommand{\meanmassfullone}{\langle \mu | \mathbf{s} \rangle_1}
\newcommand{\varmassfullone}{\sigma_{\mu | \mathbf{s},1}^2 }
\newcommand{\sigmamassfullone}{\sigma_{\mu | \mathbf{s},1} }
\newcommand{\meanmassSaone}{\langle \mu | s_a \rangle_1}
\newcommand{\varmassSaone}{\sigma_{\mu | a,1}^2 }
\newcommand{\sigmamassSaone}{\sigma_{\mu | a,1} }
\newcommand{\sigmamassSbone}{\sigma_{\mu | b,1} }
\newcommand{\varmassSbone}{\sigma_{\mu | b,1}^2 }
\newcommand{\meanSbone}{\langle s_b | s_a \rangle_1}
\newcommand{\varSbone}{\sigma^2_{b | a,1}}

\newcommand{\meanmassfulltwo}{\langle \mu | \mathbf{s} \rangle_2}
\newcommand{\varmassfulltwo}{\sigma_{\mu | \mathbf{s},2}^2 }

\newcommand{\varmassSatwo}{\sigma_{\mu | a,2}^2 }

\newcommand{\varSbtwo}{\sigma^2_{b | a,2}}

\newcommand{\nexact}{n_{\rm Tinker}}
\newcommand{\Mfiveh}{M_{500c}}
\newcommand{\Lx}{L_X}
\newcommand{\Ysz}{Y_{SZ}}
\newcommand{\Ngal}{N_{\rm gal}}
\newcommand{\maxBCG}{\hbox{\tt maxBCG}}

\newcommand{\lnS}{\ln S}

\newcommand{\ergs}{\, \mathrm{erg\ s}^{-1}}

\def\spose#1{\hbox to 0pt{#1\hss}}
\def\lta{\mathrel{\spose{\lower 3pt\hbox{$\mathchar"218$}}
     \raise 2.0pt\hbox{$\mathchar"13C$}}}
\def\gta{\mathrel{\spose{\lower 3pt\hbox{$\mathchar"218$}}
     \raise 2.0pt\hbox{$\mathchar"13E$}}}

\begin{document}

\title{A Model for Multi-property Galaxy Cluster Statistics}

\author[A.E. Evrard, P. Arnault,  D. Huterer, A. Farahi]{ August E. Evrard$^{1,2,3}$, Pablo Arnault$^{1,4}$, Dragan Huterer$^1$, Arya Farahi$^1$\\
$^{1}${Department of Physics and Michigan Center for Theoretical Physics, University of Michigan, Ann 
Arbor, MI 48109, USA.} \\
$^{2}${Department of Astronomy, University of Michigan, Ann Arbor, MI 48109, USA.}\\
$^{3}${Institut d'Astrophysique, 98bis Bd. Arago, 75012 Paris, France.}\\
$^4${\'Ecole Normale Sup\'erieure de Cachan, 61 Avenue du Pr\'esident Wilson, 94230 Cachan, France.}
}
% \affiliation{Department of Physics, University of Michigan, 450 Church St., Ann Arbor, MI  48109}
\date\today

\maketitle
\label{firstpage}

\setlength{\skip\footins}{0.6cm}

\begin{abstract}
The massive dark matter halos that host groups and clusters of galaxies have observable
properties that appear to be log-normally distributed about power-law mean scaling relations in halo mass. 
Coupling this assumption with either quadratic or cubic approximations to the mass function 
in log space, we derive closed-form expressions for the space density of halos as a function of multiple observables as well as forms for the low-order moments of properties of observable-selected samples.  Using a Tinker mass function in a $\Lambda$CDM cosmology, we show that the cubic analytic model  reproduces results obtained from direct, numerical convolution at the
$10$\% level or better over nearly the full range of observables covered by current observations
and for redshifts extending to $z=1.5$.  The model provides an efficient framework for estimating 
effects arising from selection and covariance among observable properties in survey samples.  
\end{abstract}


%\newpage

%\begin{changemargin}{-1cm}{-1cm}
%\addtolength{\textwidth}{0cm}

\section{Introduction}

Counts of galaxy clusters provide constraints on cosmological parameters 
% particularly a combination of the clustered matter density and the matter power
% spectrum amplitude 
\citep[\eg][and references therein]{Voit0410173, Allen1103.4829}, 
and test fundamental theories of gravity and cosmic acceleration 
\citep[\eg][]{Weinberg1201.2434}.  Such studies typically use
cluster samples identified via optical \citep{Rozo10}, X-ray \citep{Mantz0909.3098,
 Henry0809.3832} or thermal Sunyaev-Zel'dovich 
 \citep[SZ,][]{Benson1112.5435,  Sievers1301.0824, PlanckXX1303.5080} 
signatures of the baryons in the halos that host cluster phenomena.  These analyses
are empowered by simulation studies that calibrate the space density as a
function of halo mass, known as the \emph{mass function}, within a given 
cosmology \citep{Tinker08, Bhattacharya1005.2239, Murray1306.6721}.

Modeling the expected counts of clusters in a wide-area observational survey 
% outcomes using gas dynamic simulations to
% predict baryonic signatures is an option, the extreme dynamic range
% requirements and incomplete physics understanding preclude such an approach.
% Instead, 
requires combining the mass function with a statistical model that expresses 
the likelihood for a halo
of mass $M$ at redshift $z$ to have some intrinsic observable signal, 
$S$, detectable in the survey.  Evidence from observations \citep{Arnaud0502210, 
Maughan0703504, Pratt0809.3784, Vikhlinin0805.2207, 
Zhang1011.3018, Ruel1311.4953, Saliwanchik1312.3015, Ettori1307.7157, 
Maughan1212.0858} and simulations
\citep{Evrard08, Stanek0910.1599, Fabjan1102.2903, Munari1301.1682, 
Jiang1311.6649, LeBrun1312.5462, Biffi1401.2992} 
support a model in which the 
scaling law behavior is power-law with mass in
the mean, with approximately log-normal variance.

While scaling behavior of cluster properties has been studied for
decades \citep[see][for a recent review]{Giodini1305.3286}, most works have 
focused on correlating pairs of observed signals, $\{ S_2, S_1 \}$, or on studying 
how a single observable scales with mass, $\{S_1, M \}$.  
Simulations provide a natural environment for the latter,
since the true halo mass is known.  For observations, mass estimates 
are made indirectly from measured signals, for example
through assumption of virial or hydrostatic equilibrium, and this methodology
introduces the potential for bias and additional variance that must be
calibrated \citep[\eg][]{Rasia1201.1569, Battaglia1209.4082,  Nelson1308.6589}.  
Alternatively, masses can be inferred through inversion of a given observable-mass relation. 
In this way, an observable serves as proxy for halo mass.  

Evidence of biases in mass proxies can arise from comparisons among different 
observable signals.  
Planck satellite measurements of the thermal SZ effect in the optically-selected \maxBCG\ 
sample \citep{Planck11_optical} led to a detailed re-examination of X-ray, SZ and 
optical scaling relations by \citet{RozoI,RozoII,RozoIII}.  That study concluded that 
the Planck $\Ysz$ mass calibration was biased low by a few tens of percent, a finding
supported by independent weak gravitational lensing estimates of Planck clusters 
\citep{VonderLinden1402.2670}, although other studies are less 
supportive \citep{Israel1402.3267}.

\citet{RozoII} present a model for multivariate signal counts and other 
statistics under the assumption of a locally power-law mass function.  That model was 
employed to interpret a combined set of X-ray, SZ and optical data, resulting in a set 
of preferred scaling relations presented in \citet{RozoIII} (see their Table~4).

In this paper, we present a non-local extension of that model that expands its scope 
to effectively cover the complete dynamic range of properties displayed by the population of 
galaxy clusters.  Within a 
$\Lambda$CDM cosmology, we show that the mass function of the massive halos that host 
groups and clusters of galaxies can be represented by a low-order polynomial (in log-space) 
to a typical accuracy of a few percent, comparable to its calibrated level of precision 
from N-body simulations.  
Convolving this mass function representation with a multivariate Gaussian of logarithmic halo 
properties at fixed mass and redshift results in analytic expressions for the space density as 
a function of multiple observables and other derivative statistics.   
By offering a fast method for estimating survey sample and follow-up study outcomes, this 
formalism is intended to complement data analysis methods based on similar model 
assumptions \citep{Maughan1212.0858}.

We employ a halo mass convention of $\Mfiveh$, the mass within a spherical 
region encompassing $500$ times the critical density, $\rho_c(z)$, but the analytic 
expressions can be applied using any  choice of halo mass convention.  We use 
$\Mfiveh$ to be consistent with the scaling laws presented in \citet{RozoIII}.  
In \S\ref{sec:model}, we derive expressions for multi-observable cluster population 
statistics using low-order polynomial approximations of the mass function in log space.  
These are then applied to X-ray and SZ statistics in \S\ref{sec:applications}.  In \S\ref{sec:discussion}, 
we discuss some of the model's strengths and limitations, and we summarize our results in \S\ref{sec:summary} 

\section{Model for Low-order Moments}\label{sec:model} 

We first develop expressions for the space density of clusters as a function of observables such as temperature or luminosity, then compute first and second moments for properties of samples selected by a specific observable.   Our model expands the results presented in \citet{RozoII} and uses slightly different notation.

Consider a set of $N$ bulk observable properties $S_a$ where $a \in \{1,N \}$; these
observables can be, for example, X-ray luminosity, $L_X$, temperature, $T_X$,
gas thermal energy measured in the X-ray or SZ flux, $Y_X$ or $Y_{SZ}$, number of galaxies, $\Ngal$ or 
$\lambda$, inferred lensing mass $M_{\rm lens}$, etc. measured within some characteristic radius.   
Let 
$s_a \equiv \ln (S_a)$ represent the natural logarithms of these properties in some chosen 
basis of units (for example, $10^{44} \ergs$ for X-ray luminosity).  

We assume power-law forms for the observable--mass scaling relations.  
Choosing a halo pivot mass scale, $M_p$, at some fiducial redshift (values discussed below), and  
letting $\mu \equiv \ln (M/M_p)$, 
the vector of log-observables, $\mathbf{s}$, scales in the mean with mass as 
\begin{equation}
\langle \mathbf{s} | \mu \rangle = \boldpi + \boldalpha \mu \ ,
\label{eq:linear_fit}
\end{equation}
where the vectors $\boldpi$ and $\boldalpha$ are the normalizations and slopes of the relevant scaling laws.  We consider redshift-dependent normalizations, $\boldpi(z)$, that scale in a self-similar manner \citep{Bohringer1112.5035}.   While the slopes may also be redshift-dependent, we take them as constant here.

Individual halos drawn from the cosmic population deviate from this mean behavior in a manner that we assume is log-normal.  At all redshifts, the full probability density function, $P(\mathbf{s}|\mu)$, is described by a covariance matrix
% , $\mathbf{C}$, having
with elements $C_{ab} = \langle (s_a - \langle s_a | \mu \rangle) (s_b - \langle s_b | \mu \rangle) \rangle$, and where $C_{aa} = \sigma_a^2$ is the intrinsic log variance of the $a^{\rm th}$ observable. 
%The correlation coefficients are defined by $r_{ij} = C_{ij}/(\sigma_{s_i}
%\sigma_{s_j})$, and lie between -1 and 1.
We assume this covariance to be independent of mass and redshift.

% \subsection{Mass at fixed observable properties}
\subsection{First-order (local) mass function}

The model of  \citet{RozoII} uses a first-order Taylor expansion of the mass function, $n(\ln M)$ (with dimension of number density per $\ln (M)$), around some pivot mass $M_p$, 
\begin{equation}
\none(\mu) = A e^{-\betaone \mu} \ ,
\label{eq:none}
\end{equation}
where $A$  and $\betaone$ are the local amplitude and (negative) slope of the mass function evaluated at the pivot, $\mu = 0$.   Note that $A$ and $\betaone$ are functions of redshift, cosmology, and pivot location,  as explained in \S 3 below.  The subscript on the space density indicates the first-order nature of the mass function expansion.
%and the both have positive definite because the mass function in all viable cosmological models monotonically declines with mass.  
%  = n_{\rm exact}(\mu = 0)$, and $\betaone = -[(d/d\mu) \ln n_{\rm exact}](\mu = 0) $. Note that $\betaone(M) > 0$ for all $M$.

Using equations~(\ref{eq:none}) and (\ref{eq:linear_fit}), Bayes theorem in the form 
$P(\mu|\mathbf{s}) = P(\mathbf{s}|\mu) P(\mu) / P(\mathbf{s})$ 
allows us obtain the mean and variance of the log mass selected by a fixed combination of observables.  
% $P(\mu | \mathbf{s})$ 
% \end{equation}
% 
In the first-order approximation to the mass function, this probability 
is Gaussian, with mean and variance
\begin{eqnarray}
\label{eq:meanmass1} 
% \langle \mu | \mathbf{s} \rangle_1 
\meanmassfullone &=&  \left[ \boldalpha^T \mathbf{C}^{-1} (\mathbf{s} - \boldpi) \,
 - \betaone \right] \sigma_{\mu | \mathbf{s},1}^2 \ , \\
 % [0.1cm]
\label{eq:varmass1} 
% \sigma_{\mu | \mathbf{s},1}^2 
\varmassfullone &=& (\boldalpha^T \mathbf{C}^{-1} \boldalpha )^{-1}. 
\end{eqnarray}

% Eq \ref{eq:meanmass1} and \ref{eq:varmass1}

In the case of a single observable quantity $s_a$, these expressions reduce to
% \begin{eqnarray}
% \label{meanmassSa}
$ \meanmassSaone = (s_a-\pi_a) / \alpha_a - \betaone  \varmassSaone $
% [0.1cm]
% \label{sigma_simple}
and 
% \sigma_{\mu}^2 \equiv \sigma_{(\mu | s) \betaone}^2 
$\varmassSaone =  \left( \sigma_a / \alpha_a \right)^2$.
% \end{eqnarray}
The mean mass is biased low relative to the assumed scaling by an amount given by the product of the local slope of the mass function and the mass variance of the chosen observable.  
% The latter is independent of mass and redshift in our model.  

%\dragan{DON'T NEED:}
%while, in the case of two properties $s_1$ and $s_2$, we have (introducing
%$u_i \equiv s_i - a_i$),
%\begin{widetext}%% enables single-column output 
%\begin{eqnarray}
%\label{mean2} 
%\langle \mu | s_1, s_2 \rangle_{\betaone} 
%& = & \frac{\sigma_{\mu_1}^{-2} \alpha_1^{-1} u_1 + \sigma_{\mu_2}^{-2}
%  \alpha_2^{-1} u_2 - r \sigma_{\mu_1}^{-1} \sigma_{\mu_2}^{-1} [\alpha_2^{-1}
%    u_2 + \alpha_1^{-1} u_1] }{\sigma_{\mu_1}^{-2} + \sigma_{\mu_2}^{-2} - 
%2r\sigma_{\mu_1}^{-1}\sigma_{\mu_2}^{-1}} -  \betaone \sigma_{(\mu | s_1, 
%s_2)\betaone}^2 \\ \nonumber \\[0.0cm]
%\label{variance2} \sigma_{(\mu | s_1, s_2)\betaone}^2 
%& = & \frac{1-r^2}{\sigma_{\mu_1}^{-2} + \sigma_{\mu_2}^{-2} - 
%  2r\sigma_{\mu_1}^{-1}\sigma_{\mu_2}^{-1}  } \ . 
%\end{eqnarray}
%\end{widetext}

% \subsection{Halo number density as a function of observable properties}
\medskip
\noindent
{\bf Space density of multiple observables.}  Convolving equation~(\ref{eq:none}) with the log-normal likelihood, $P(\mathbf{s}|\mu)$, yields the halo number density as a function of the full vector of observable properties, 
\begin{eqnarray}
\nonumber
& \none(\mathbf{s}) =  A_1^\prime
\exp \left[ - \frac{1}{2} \left( (\mathbf{s} - \boldpi)^T \mathbf{C}^{-1} (\mathbf{s} - \boldpi) 
 - \frac{\meanmassfullone^2}{ \varmassfullone} \right) \right],  \\ 
  & A_1^\prime =  A \ \sigmamassfullone \,  \left( (2\pi)^{N-1} |\mathbf{C}| \right)^{-1/2}  .~~~~~~~~~~~~~~~~~~~~~
\label{eq:nonefull}
\end{eqnarray}

% In the case of a single property, this expression reduces to
% \begin{equation} \label{n1}
% n_{\betaone}(s) = \frac{A}{\alpha} \exp\left[- \betaone \left( \frac{s-a}{\alpha} - \frac{1}
% {2}\betaone \sigma_{\mu}^2 \right)\right] \ .
% \end{equation}

% \subsection{Selecting on some observable property}
\medskip
\noindent
{\bf Observable-selected samples.}  
Now consider selecting a sample by a certain observable, $s_a$.  With the full space density 
above, we can derive the probability density function (PDF) of any second observable, 
$s_b$, by $P(s_b | s_a) = n(s_a, s_b)/n(s_a)$.  The result is also  Gaussian with mean and variance 
\begin{eqnarray}
\label{eq:meansb1}  
%   \langle s_2 | s_1 \rangle_{\betaone} 
\meanSbone &=&   \pi_b + \alpha_b [ \, \meanmassSaone + \betaone \, \rab \, \sigmamassSaone \, \sigmamassSbone \, ] \, , \\
\label{eq:varsb1}  
% \sigma_{(s_2 | s_1) \betaone}^2 
\varSbone &=& \alpha_b^2\, [ \, \sigmamassSaone^{2} + \sigmamassSbone^{2} - 2\rab \,  \sigmamassSaone\sigmamassSbone \, ] \, , 
\end{eqnarray}
where $\rab$ is the correlation coefficient between properties $s_a$ and $s_b$ at fixed mass.  When the intrinsic correlation of these observables is non-zero, then a shift in the mean of $s_b$ is induced that is similar in form to the bias of equation~(\ref{eq:meanmass1}) but with opposite sign if $\rab$ is positive.  This effect can be understood by the fact that the dominant lower mass halos that scatter upward into the chosen $s_a$ bin will also have a positive deviation from the mean $s_b$ if $\rab$ is positive.  If $\rab$ is negative, the effect is reversed.  

Along with these purely observable properties, one can compute the 
correlation coefficient between mass and $s_b$ for samples chosen by $s_a$, finding 
\begin{equation} 
\label{eq:rmub1}
r_{(\mu b | a), 1} = 
\frac{\sigmamassSaone /  \sigmamassSbone  - \rab}{[1 - \rab^2 + ( \sigmamassSaone / \sigmamassSbone - \rab)^2]^{1/2}} \ .
\end{equation}
In the case of uncorrelated observables ($\rab = 0$), a positive correlation between mass and $s_b$ is induced by the fact that halos with lower mass that scattered up into the $s_a$ bin will also have lower $s_b$, and vice-versa.  In the limit that the selection property, $s_a$, is a much better mass proxy than $s_b$, such that $\sigmamassSaone /  \sigmamassSbone \simeq 0$, then the correlation of $s_b$ and mass takes on the opposite sign of the intrinsic correlation, $\rab$.  Relative to the mean behavior in the selection bin, halos with of lower mass, $\Delta \mu < 0$, will have positively enhanced selection signal at that mass, $\Delta s_a > 0$, and then $\Delta s_b \simeq  \rab  \Delta s_a \simeq - \rab  \Delta \mu$.

%%%%%%%%%%%%%%%%%%%%%%%%%%%%%%%%%

\subsection{Higher-order (non-local) mass functions} 

The first-order model functions well over a relatively narrow range in mass or observable near the chosen pivot point.  We now wish to extend the range of the model by introducing quadratic and cubic terms into the mass function approximation.  We derive here exact expressions for the quadratic case, and approximate expressions for the cubic case, and show below that the latter are accurate to better than $10\%$ for a wide range of halo mass scales and redshifts.  
% covering the regime probed by current and near future cluster surveys.  

The second-order model uses a mass function, 
\begin{equation}
\ntwo(\mu) = A e^{-\betaone \mu - \frac{1}{2} \betatwo \mu^2} \ ,
\label{eq:ntwo} 
\end{equation}
where $\betatwo$ is the magnitude of the second derivative of the mass function at the pivot mass scale, which is negative for massive halos in \LCDM\ cosmologies.

The convolution remains analytic, and the halo number density as a function of multiple observables retains the form of equation~(\ref{eq:nonefull}), 
% \begin{equation}
% \ntwo(\mathbf{s}) = A_2^\prime \, \exp \left[ - \frac{1}{2} \left( (\mathbf{s}
%  - \boldpi)^T \mathbf{C}^{-1} (\mathbf{s} - \boldpi) - 
%  x_{\mathbf{s}} \frac{\meanmassfullone^2}{ \varmassfullone} \right) \right] , 
% \label{eq:ntwofull} 
% \end{equation}
% where $A_2^\prime = \sqrt{x_s} A_1^\prime$ and
but adding local curvature reduces the weight of lower-mass halos scattered upward into the signal bin. 
Consequently, the gaussian distribution of halo mass at fixed observable properties has a compressed mean and variance relative to the first-order treatment, 
\begin{eqnarray}
\label{eq:meanmass2} 
\meanmassfulltwo &=& x_{\mathbf{s}} \, \meanmassfullone \, , \\
\label{eq:varmass2} 
\varmassfulltwo  &=&  x_{\mathbf{s}} \,  \varmassfullone \, ,
\end{eqnarray}
\noindent
where the compression factor,  $x_{\mathbf{s}} \equiv (1 + \betatwo \, \varmassfullone)^{-1}$, is less than unity and is well approximated by $1 - \betatwo \varmassfullone$ for most of the applications discussed below.  

% In the case of a single observable quantity, this expression reduces to
% \begin{equation} \label{n2}
% \ntwo(s_a)  =  \frac{A}{\alpha_a} \sqrt{x_a} \ \exp \left[ -\frac{1}{2} \left( \frac{(s_a-\pi_a)^2}
% {\sigma_a^2} -  x_a \frac{ \langle \mu | s \rangle_1^2}{\sigma_{\mu | a,1}^2} \right) \right] \ .
% \end{equation}

Consider again the case of two observables $s_a$ and $s_b$. 
The PDF of having observable $s_b$ in a population selected by observable $s_a$ also 
remains Gaussian, and the expressions written in mass equivalents are somewhat simpler.  
Letting $\delta_i = (s_i - \pi_i ) / \alpha_i$, then the mean and variance for the second-order 
mass function approximation are  
\begin{eqnarray}
\label{eq:meansb2}  
\langle \delta_b | s_a \rangle_2 &= x_a \left[ \meanmassSaone + 
(\betaone + \betatwo \delta_a) \,  \rab \,  \sigmamassSaone \, \sigmamassSbone   \right]  , \\
\label{eq:varsb2}  
\frac{ \varSbtwo } { \alpha_b^2 }  &= x_a  \left[  \frac{ \varSbone } { \alpha_b^2 }  
+ \betatwo  \, \varmassSaone \, \varmassSbone \, (1-\rab^2) \right]  .  ~~~~~~
\end{eqnarray}
The first expression indicates that the mean observable now senses the curvature in the mass 
function through the $ \betatwo \delta_a$ term, where recall that $\delta_a$ is measuring 
the equivalent log-mass distance from the pivot location.  

In the limit of uncorrelated 
observables ($\rab = 0$), the second expression reduces to 
$x_a \varmassSaone + \varmassSbone$, as it should 
since the mass function curvature affects the mass variance in the selection variable but not 
that of the non-selection variable.  

Finally, the correlation coefficient between mass and property $s_b$ at fixed $s_a$ is now given by
\begin{equation}
\label{eq:rmub2} 
r_{(\mu b|a),2} =  \frac{  \sigmamassSaone / \sigmamassSbone  - \rab}
{ [(1 - \rab^2) / x_a  + (  \sigmamassSaone / \sigmamassSbone  - \rab)^2]^{1/2}}  .
\end{equation}
For uncorrelated observables this expression again reduces to equation~(\ref{eq:rmub1}) with $\varmassSatwo$ replacing $\varmassSaone$.  

In the Appendix, we show a further extension to the third-order,
with $\nthree(\mu)= A e^{-\betaone \mu - \frac{1}{2}\betatwo \mu^2 - \frac{1}{6} \betathree \mu^3}$.

%%%%%%%%%%%%%%%%%%%%%%%%%%%%%%%%%%%%%%%%%%%%%%%%%%%%%%
\section{Applications to Observable Cluster Properties}\label{sec:applications} 

We now evaluate the utility of the above expressions by comparing their predictions to expectations
calculated via explicit local convolution of the Tinker mass function \citep{Tinker08}.   For hot gas 
observables, we examine the X-ray luminosity at soft photon energies, $\Lx$ and the total gas thermal 
energy as determined by the thermal SZ, $\Ysz$.  These cases represent examples of relatively high and
low-scatter mass proxies, respectively.   We also examine the case optical richness, $\Ngal$, a relatively
high-scatter proxy for which the correlation with hot gas properties at fixed mass is currently poorly understood.   

We perform analysis at $z=0.23$, the redshift where local mass-observable relations used 
here are calibrated, and also at $z=1.5$.  The higher redshift is chosen to be representative 
of the outer reaches of near-term cluster surveys and is also an epoch at which the mass 
function is both steeper and more strongly curved compared to low redshift, aspects that 
make the higher-order corrections more important.  

\subsection{Mass function and log-space polynomial fits}\label{sec:tinkerMF}

The Tinker mass function employs an updated version of the normalized mass fraction functional, $f(\sigma)$, in the form originally calibrated for CDM cosmologies by \citet{Sheth9907024} and \citet{Jenkins0005260}, 
\begin{equation}
\nexact(\mu,z) = \frac{dn}{d \mu} = \frac{\overline{\rho}_m(z)}{M} \frac{d \ln \sigma^{-1}(M,z)}{d \mu} f(\sigma) \ ,
\label{eq:tinkerMF}
\end{equation}
where $\overline{\rho}_m(z)$ is the mean cosmic matter density, and $\sigma^2(M,z)$
is the linearly evolved variance of matter density fluctuations filtered on a mass scale 
$M \equiv M_p e^\mu$, both 
evaluated at redshift $z$.  In this work, we use the mass function 
calculator
% \footnote{http://hmf.icrar.org/} 
published by \citet{Murray1306.6721} 
and employ the Tinker fit for $f(\sigma)$ \citep{Tinker08} using the CAMB transfer function of a WMAP7
cosmology.   The cosmological parameters for the $\Lambda$CDM model are: scaled Hubble constant, 
$h=0.704$, baryon, cold dark matter and dark energy parameters of, $\Omega_{\rm b} = 0.0455$, 
$\Omega_{\rm c} = 0.226$,  $\Omega_{\rm DE} =  0.728$, respectively, spectral index, $n_s=0.967$, 
and present amplitude of matter density fluctuations, $ \sigma_8 = 0.81$.

At our two fiducial redshifts, we compute the coefficients $A$, $\betaone$, $\betatwo$ and $\betathree$ 
by taking numerical derivatives of $\nexact(\mu,z)$ at the chosen pivot mass.  Values of the pivot mass, 
$M_p$, in units of $10^{14} \msol$ along with the fit parameters are shown in 
Table~\ref{tab:betaparams}.   At higher redshifts, the mass function steepens and becomes 
more strongly curved.   For observables with large mass variance the second-order 
compression factor, $1+\betatwo \sigma_{\mu|s}^2$ can be important since 
$\betatwo =0.70$ and $1.22$ at redshifts $z=0.23$ and $1.5$, respectively.

\begin{table}
\centering
%\begin{minipage}{90mm}
\caption{Mass function expansion parameters at two redshifts for a WMAP7 cosmology.  The pivot mass, $M_p$, is in units of $10^{14} \msol$ and the amplitude, $A$, is in units of $10^{-6} \, {\rm Mpc}^{-3}$.  }
\begin{tabular}{cccccc}
$z$ & $M_p$  & $A$  & $\betaone$ & $\betatwo$ & $\betathree$ \\
\hline
0.23 & 2.0 & $1.944$  & 1.97  & 0.70   & 0.40 \\
1.5  & 1.0  & $0.293$ & 3.07  & 1.20  & 0.73  \\
\hline
\end{tabular}
%\end{minipage}
\label{tab:betaparams}
\end{table}

\subsection{Scalings for $\Lx$, $\Ysz$, and $\Ngal$}\label{sec:scalings}

As specific examples of scaling laws we use results derived in the pan-chromatic study of \citet{RozoIII}.  
In that work, $L_X$ is the X-ray luminosity in the rest-frame $[0.1,2.4]$ keV band, 
expressed in units of $10^{44} \mathrm{ergs/s}$ and measured within a cylindrical 
aperture of radius $R_{\rm 500c}$.  The thermal SZ signal, 
$D_{A}^2 Y_{SZ}$ where $D_A$ is the angular distance of the source, is given in units of
$10^{-5} \mathrm{Mpc} ^2$ and is measured within the spherical aperture of radius 
$R_{\rm 500c}$.  The optical richness, $\Ngal$, is the count of red sequence galaxies 
determined by the \maxBCG\ cluster-finding algorithm within an estimated sphere 
of radius $R_{\rm 200c}$ \citep{Koester0701265}.

We use the observable---mass parameters, $\pi_s$, $\alpha_s$, and 
$\sigma_{\ln S}$, at $z = 0.23$ given in Table 4 of \citet{RozoIII}.  
The parameters, with normalizations rescaled to our choice 
of pivot mass, $2 \times 10^{14} \msol$, are summarized in Table~\ref{tab:rozoparams}.  
 At $z=1.5$, we employ self-similar scalings for normalizations of $\Lx$ and $\Ysz$ discussed below.  
  
For the optical richness, we use a simple inversion of the scaling of lensing mass at fixed $\Ngal$ and consider both the published value of the scatter as well as a more optimistic value that is appropriate 
for the multi-color richness estimator, $\lambda$ \citep{Rykoff1104.2089, Rozo1303.3373}.   
The improvement of a factor of four in variance has significant implications that we illustrate below.  

% Recall that $s = \ln(S/S_p)$, which can be denoted $\lnS$, with $S$ implicitly given in units of the pivot value, $S_p$. Below, we employ the latter notation, $\lnS$, in order to be explicit about when logarithms of the observables are being used. 

\begin{table}
\centering
%\begin{minipage}{90mm}
\caption{Observable-mass scaling parameters at $z=0.23$ and pivot mass, $M_p = 2 \times 10^{14} \msol$, from \citet{RozoIII}. See text for unit definitions.  Along with the published scatter values for $\Ngal$, we also consider a smaller value in parentheses based on the $\lambda$ richness estimator \citep{Rykoff1104.2089}.  }
\begin{tabular}{ccccc}
 $S$ & $S_p = e^{\pi_s}$ & $\alpha_s$ & $\sigma_{\lnS}$ & $\sigma_{\mu|s}$  \\
 \hline
 \noalign{\vspace{2.0truept}}
$\Lx$ & 0.61 & 1.55 & 0.39 &  0.252 \\
$D_A^2 \Ysz$ & 0.62  & 1.71 & 0.15 &  0.088 \\
$\Ngal$ & 37  & 0.94 & 0.42 (0.21) &  0.45 (0.23) \\
\noalign{\vspace{2.0truept}}
\hline
\end{tabular}
%\end{minipage}
\label{tab:rozoparams}
\end{table}

% As we  test how accurately our formulae approximate those 
% derived from the Tinker mass function in the following sections, 
% we employ the fractional difference
% \begin{equation}
% \delta(Q) = \frac{Q - Q_{\rm Tinker}}{Q_{\rm Tinker}} \ ,
% \end{equation}
% for some linear measure, $Q$.  Below, we test cases 
% of $Q = n_i(\lnS)$, $\langle S_b|S_a \rangle$, and $\sigma_{(b|a),i}$, 
% where $i = 1, 2, 3$ is the order of the mass function approximation.  

\subsection{Application to cluster number counts}\label{sec:counts}

We first compare the approximate analytical formulae, $n_i(\ln S)$, to
$n_{\rm Tinker}(\ln S)$, where the latter quantity is obtained by 
explicit convolution of the Tinker mass function, 
$\int n_{\rm Tinker}(\mu) P(\ln S|\mu) d\mu$.

\begin{figure}%[htbp]
\begin{center}
\includegraphics[scale=0.45]{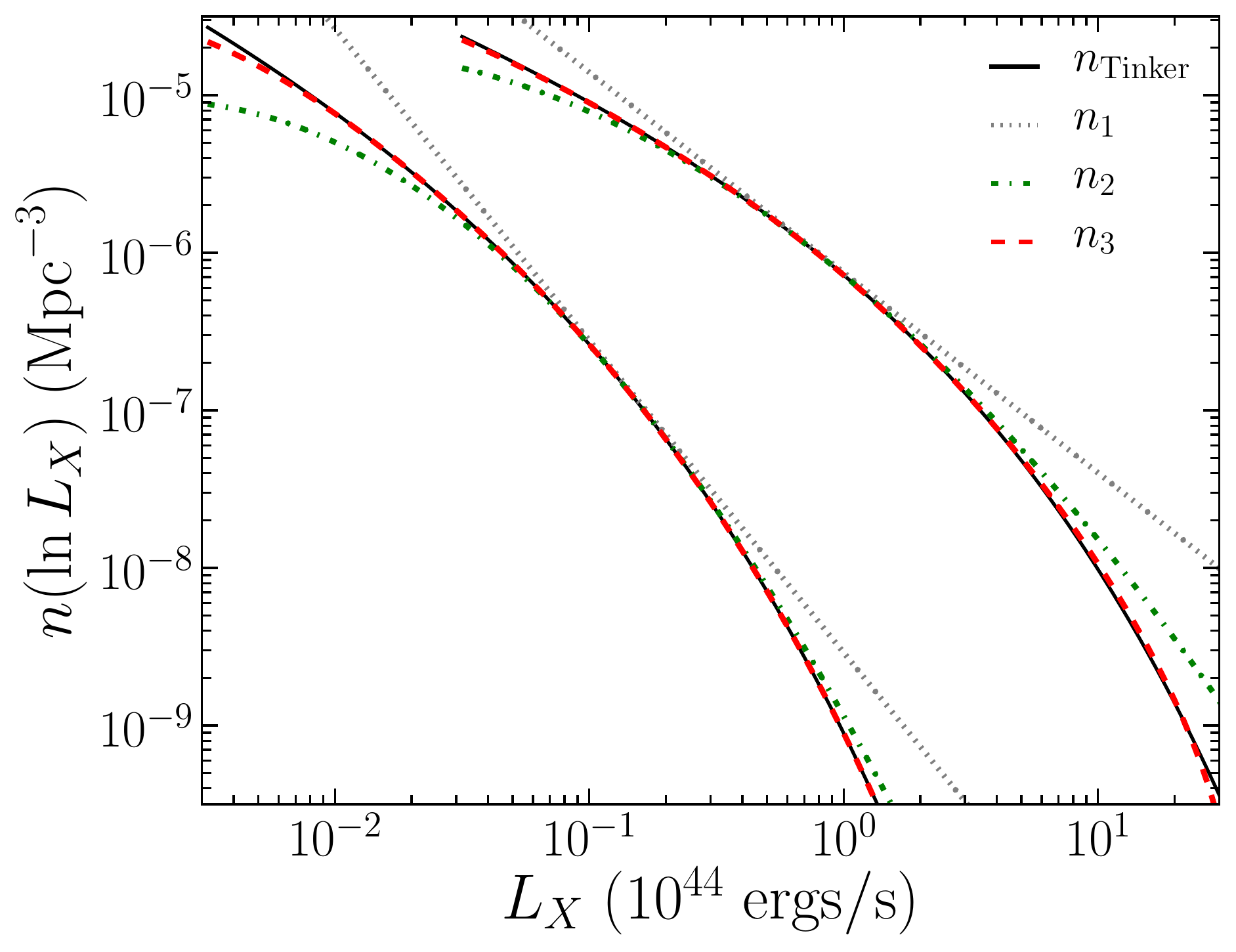}
\vspace{-0.9cm}
\end{center}
\caption{Comoving space density of halos as a function of $L_X$  evaluated at 
redshift $z = 0.23$ (upper curves) and $z=1.5$ (lower), 
  assuming self-similar redshift evolution of the X-ray normalization for the latter.  
  Black lines show expectations from 
  local convolution of the Tinker mass function, while the dotted (grey), 
  dot-dashed (green) and dashed (red) curves show
  our analytic expressions based on first, second and third-order mass function 
  expansions, respectively.  
  Pivot masses of $M_p = 2$ and $1 \times 10^{14} M_{\odot}$ are used at $z = 0.23$  and $1.5$, 
   respectively, and the latter are offset by a factor of $0.1$ in $\Lx$ for clarity.
   Deviations of the approximations from the direct Tinker convolution are plotted 
  in Figure~\ref{fig:diffs_ns}.    
  }
\label{fig:n_Lx}
\end{figure}

In Figure~\ref{fig:n_Lx} we show the X-ray luminosity function at redshifts 
$z = 0.23$ and $z=1.5$, derived using the parameters in Tables~\ref{tab:betaparams} 
and \ref{tab:rozoparams}.  The normalization of the high redshift $\Lx$--$M$ relation 
is scaled using the  assumption that the soft-band luminosity at fixed mass scale in 
a self-similar fashion, $L(z) \sim H(z)^2$ \citep{Bohringer1112.5035}.   While the 
first-order model is locally accurate, the second- and third-order models extend 
the accuracy over increasingly wider ranges X-ray luminosity.  The third-order model traces 
well the local Tinker convolution results across more than two decades in luminosity.  

\begin{figure*}%[t]
%\begin{center}
%\includegraphics[width=3.3cm]{legende_difference}
%\end{center}

\includegraphics[height=6.7cm, width=8.3cm]{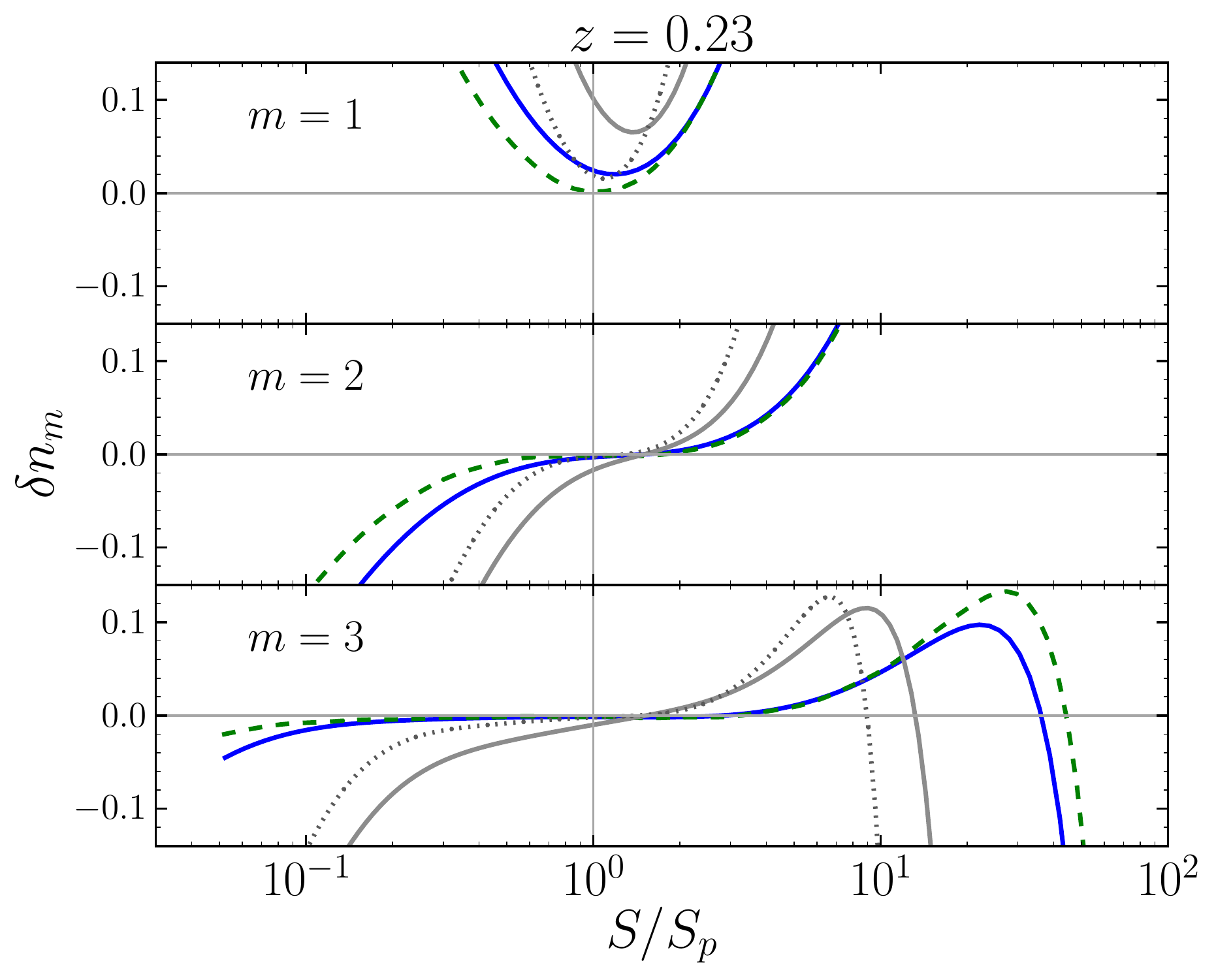}
\ \ \ \ \ \ \includegraphics[height=6.7cm, width=8.3cm]{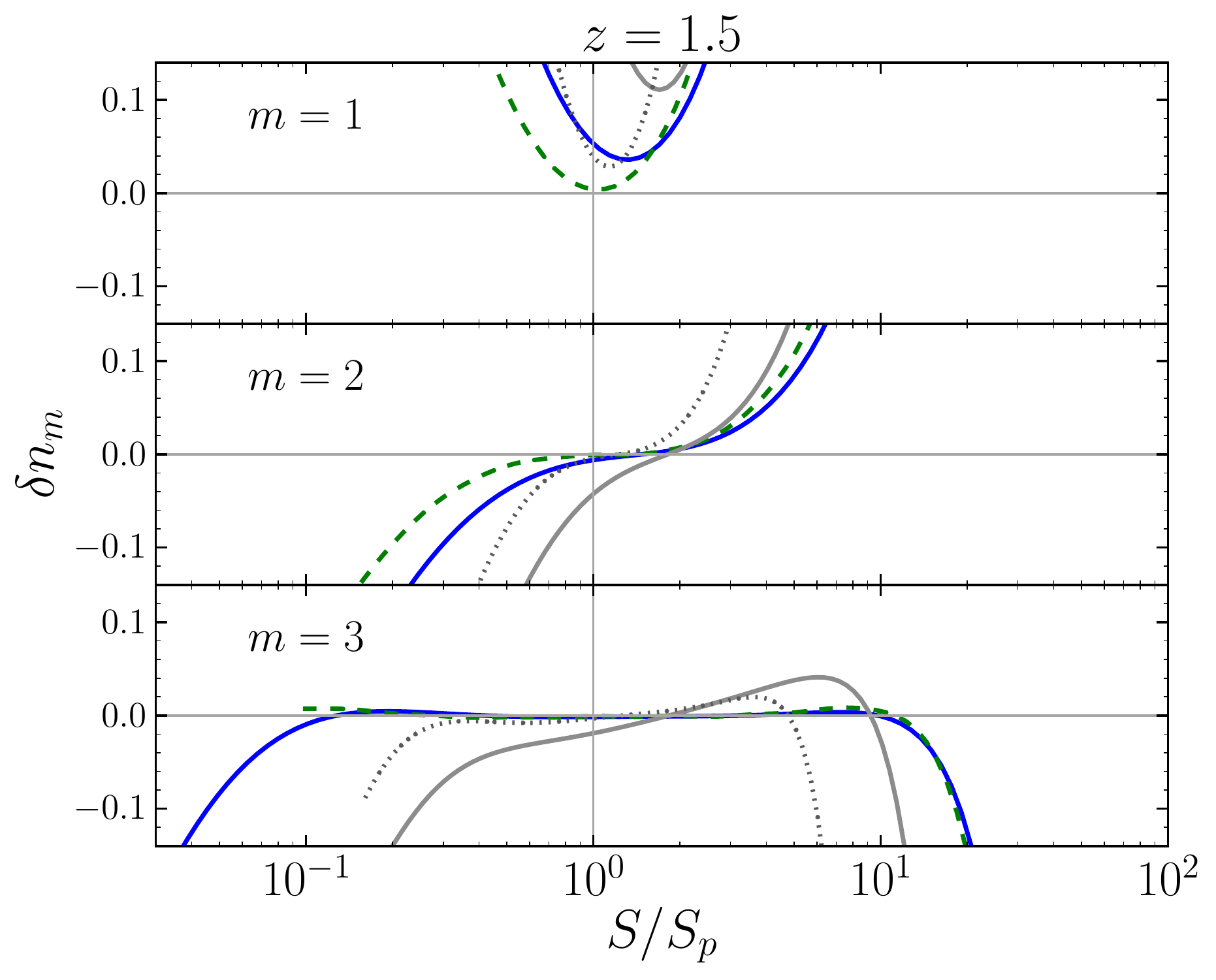}
\vspace{-0.3cm}
\caption{Fractional error in number counts, $\delta n_m(\ln S) \equiv n_m(\ln S) / n_{\rm Tinker}(\ln S) - 1$, are shown for first, second, and third-order expansions of the log-space mass function (top to bottom) at $z=0.23$ (left) and $1.5$ (right).   The observable
  signals $S$ represented are X-ray luminosity, $L_X$ (blue solid lines), the SZ
  flux, $D_A^2 Y_{SZ}$ (green dashed) and the optical richness, $\Ngal$, using 
  either the high scatter value, $\sigma_\mu = 0.45$ (grey solid) or low value, $\sigma_\mu = 0.23$ 
  (grey dotted).   Signals are plotted relative to the log-mean values, $S_p$, expected at the pivot 
  mass scales of $2$ and $1 \times 10^{14} \msol$ (left and right, respectively) using 
  parameters in Table~\ref{tab:rozoparams}.   
  }
  \label{fig:diffs_ns}
\end{figure*}

In Figure~\ref{fig:diffs_ns}, we evaluate the accuracy of the different orders 
for all observables listed in Table~\ref{tab:rozoparams}.   Values of the properties range from 
$0.03-100$ for $\Lx/10^{44} \ergs$ and $D_A^2 \Ysz / 10^{-5} \, {\rm Mpc}^2$ and $3-1000$ 
for $\Ngal$.  

At first order, the first-order counts always overestimate the Tinker convolved estimates.  The 
normalization error is approximately $1-\sqrt{x_s}  \simeq \betatwo \varmassSaone/2$.  
The first-order approximation is thus most accurate for $\Ysz$ at low redshift and 
worst for $\Ngal$ at high redshift.   With $x_s = 0.80$, the first-order $\Ngal$ 
counts at $z=1.5$ lie more than $10\%$ above the Tinker convolution, 
even near the pivot point.  When the optical richness scatter is reduced by a factor of
 two (dotted grey lines), the first-order  improve dramatically.  
 In all cases, the range within which the first-order estimates are accurate 
 is limited to a factor of a few close to the pivot location.  

At second order, the dynamic range over which the counts lie within $5\%$ of the Tinker expectations widens, reaching nearly a decade for $\Ysz$ and $\Lx$ at low redshift.  The approximation is always better at low redshifts where the second and third derivatives, $\betatwo$ and $\betathree$, of the mass function are lower.  Note that the zero crossing is shifted upward from the pivot location by an amount that scales with the signal variance.  
 
The third-order approximation is accurate to within $5\%$ in number over more than two 
decades in both $\Lx$ and $\Ysz$, with the larger errors only occurring at high 
signal values that correspond to $M_{500}$ masses above $1$ and $0.5  \times 10^{15} \msol$ 
at low and high redshift, respectively.   Such massive halos are quite rare, with 
space densities of roughly $20$ and $0.2$ per cubic gigaparsec at $z=0.23$ and $1.5$.  
For statistical cluster samples that typically reach space densities many times higher, 
the third-order model is capable of yielding highly accurate estimates of counts 
as a function of observable properties.  

One can always vary the pivot mass scale in order to improve the quality fit over a certain range of signal.  However, even with a tuned pivot, the first and second-order models cannot reach $10\%$ accuracy over the wide dynamic range of observables covered by current surveys.  In particular, unless the pivot mass scale is chose to be very high, it is very difficult for these models to obtain good accuracy at large masses and high redshifts.

\subsection{Observable-selected sample expectations}

% Surveys select clusters according to some observed property.  
Dedicated follow-up observations or joint studies of overlapping surveys 
at different wavelengths can allow multiple properties to be measured for 
clusters selected by a particular observable.  
Here, we explore expectations for such secondary properties 
based on the different orders of the multivariate model of \S\ref{sec:model}.  

\begin{figure}% [t]
\begin{center}
\includegraphics[scale=0.45]{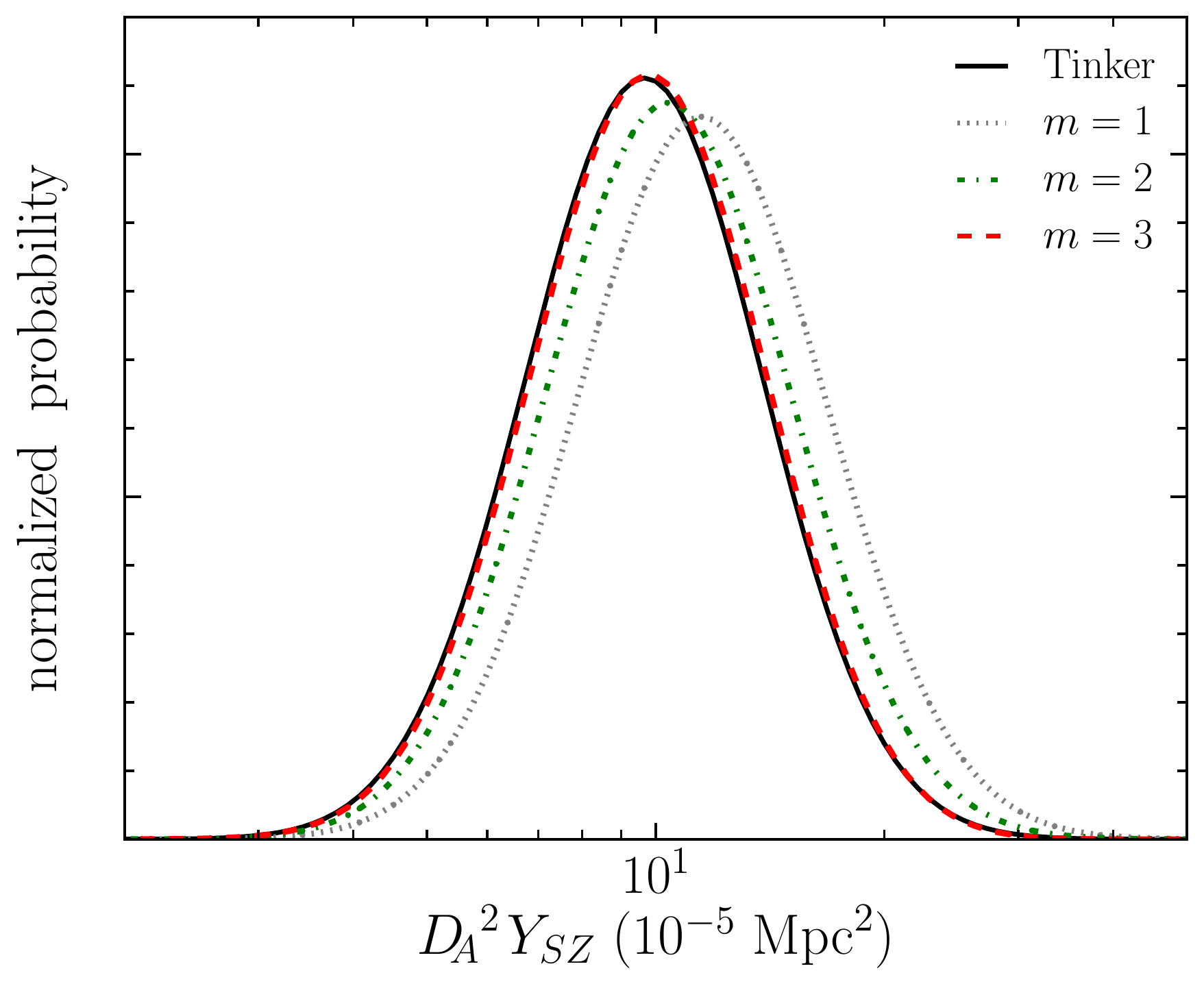} 
\end{center}
\vspace{-0.5cm}
\caption{ Probability density function, $P(\ln S_b|S_a)$, for the SZ signal, 
$S_b = D_A^2 Y_{SZ}$, at a chosen X-ray luminosity, 
$S_a = L_X = 10^{45} \mathrm{ergs/s}$ at $z=0.23$ as inferred from direct convolution of 
the Tinker mass function (black line) or from the approximate model using first (grey, dotted), 
second (green, dot-dashed) and third-order (red, dashed) forms.   
}
\label{fig:condProbExample}
\end{figure}

% Given a selection property, $a$, the conditional probability of observing another property, $b$, is given by the ratio of space densities, $P(s_b | s_a) = n(s_b, s_a) / n(s_a)$.  
Figure~\ref{fig:condProbExample} shows an example for low redshift ($z=0.23$) in the form 
of the SZ thermal decrement expected for clusters selected at fixed X-ray luminosity.  
We chose a relatively bright luminosity of 
$L_X =  10^{45} \mathrm{ergs/s}$ appropriate for a mass scale of $1.2 \times 10^{15} \msol$,
 a factor of six above the low redshift pivot point.   The correlation coefficient, $r$, 
between $\ln (D_A^2 Y_{SZ})$ and $\ln (L_X)$ is expected to be positive since both signals scale as 
positive powers of the intracluster gas mass and temperature \citep{Stanek0910.1599, Angulo1203.3216}.  
We take $r=0.5$ as an example value.  

In Figure~\ref{fig:condProbExample}, the first order likelihood overestimates the mean by 
$\sim \! 20\%$.  This relatively small error arises from near-cancellation of the much larger errors 
made in both $n_1(\ln Lx, \ln D_A^2 \Ysz)$ and $n_1( \ln \Lx)$ relative to their Tinker values at 
this high mass.  The first-order model is surprisingly accurate for conditional likelihoods.  

By sensing the local mass function curvature, the second-order estimate, 
equation~(\ref{eq:meansb2}), corrects the first-order logarithmic mean by 
roughly $\alpha_b (x_a - 1) \simeq -0.07$, which reduces, but does not fully eliminate, 
the discrepancy with the local Tinker convolution.  The third-order succeeds in 
matching the local Tinker probability with high accuracy.   

The behavior of the low-order analytical likelihoods depends primarily on 
the mass variance of the selection variable and the signal value relative to the pivot location.  
Choosing $\Ysz$ as the selection variable yields a compression factor, $x_s$, of $0.995$, much 
closer to unity than the $0.957$ value for $\Lx$, and thus the errors at all orders are considerably 
reduced.   

Keeping $\Lx$ as the selection variable but selecting at an X-ray luminosity {\sl below}, rather than 
above,  the pivot luminosity, $L_p$  results in different behavior because the corrections at odd and 
even orders have different signs.  Selecting at $\Lx = L_p/A$, where $A>1$, the first-order mean 
$\ln D_A^2 \Ysz$ lies lower than the Tinker expectation, but the overshoot is smaller in 
magnitude than for the case of selecting at $\Lx = A L_p$.  The second-order correction 
then slightly overcorrects,  with the mean {\sl above} the Tinker estimate.  The third-order 
applies a small, negative correction to closely align with the Tinker value.

\subsection{Covariance between high-scatter mass proxies }

As a final demonstration of the model, we examine effects of covariance between 
$\Ngal$ and $\Lx$, the relatively poor mass proxies used here with 
$\ln M$ scatter of $0.45$ and $0.25$, respectively (see Table~\ref{tab:rozoparams}).  
There are currently no theoretical or empirical 
constraints on the covariance between cold and hot baryons at fixed halo mass and redshift, 
so there is complete freedom in choosing the correlation coefficient, $r$, linking their deviations
about the mean scaling behaviors.   We then contrast the outcome of this case 
against an example using the improved richness estimator, $\lambda$, 
assumed to have mass scatter of $0.23$, for halos selected using the superior mass proxy, 
$\Ysz$, with mass scatter of $0.088$.  

\begin{figure}%[htbp]
\begin{center}
\includegraphics[scale=0.45]{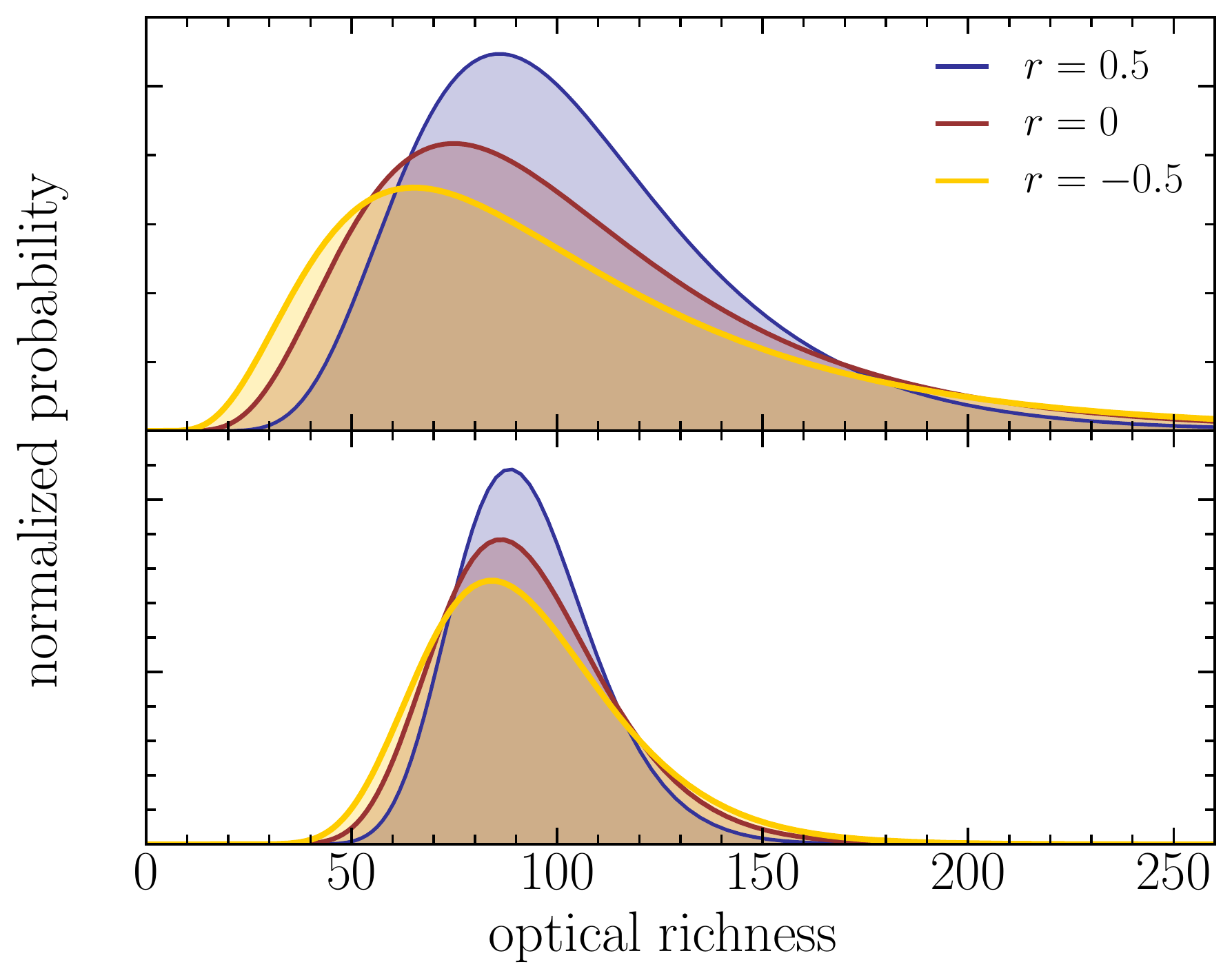} 
\end{center}
\vspace{-0.5cm}
\caption{ Expectations for optical richness at $z=0.23$ are demonstrated for two extreme cases.  
The top panel selects halos with $\Lx = 2.5 \times 10^{44} \ergs$ and assumes a log-mass scatter 
at fixed richness of $0.45$, appropriate for the original \maxBCG\ sample  $\Ngal$ richness estimator.  
The lower panel selects halos with $D_A^2  \Ysz = 3 \times 10^{-5} \, {\rm Mpc}^2$ and 
assumes a log-mass scatter at fixed richness of $0.23$, appropriate for the improved 
$\lambda$ richness estimator.  In both panels, the different PDF's arise from different 
choices of correlation coefficient between richness and the selection variable, 
as indicated by the legend.   The third-order, log-space approximation to the mass function 
is used in all calculations.  
% In comparison to the low-scatter case of the lower panel, the poorer 
% mass proxy selection observable ($\Lx$) and larger 
% $\Ngal$--$M$ scatter combine to produce a much wider 
% dynamic range in optical richness.  
}
  \label{fig:NgalGivenLxYsz}
\end{figure}

From equation~(\ref{eq:meansb2}), with the aforementioned values for the mass scatter for the 
two proxies, the shift in mean expected $\ln\Ngal$ at fixed $\Lx$ as $r$ is varied from $-0.5$ to $0.5$ 
will be of order $0.11$ times the magnitude of the local logarithmic slope of the mass function.  At a 
mass scale of $5 \times 10^{14} \msol$, the latter is of order three, implying a shift of roughly 30\%.  
In addition, the variance, 
equations~(\ref{eq:varsb1}) and (\ref{eq:varsb2}), is maximized when the properties strongly 
anti-correlate while it is minimized as $r \rightarrow 1$.  

In the top panel of Figure~\ref{fig:NgalGivenLxYsz}, we show the likelihood of optical richness at 
$z=0.23$ for halos selected to have an X-ray luminosity of  $2.5 \times 10^{44} \ergs$, 
which highlights a mass $M_{500} \simeq 5 \times 10^{14} \msol$.  The third-order 
estimator is used to 
calculate the conditional likelihood, and we show expectations for three discrete values of 
the correlation coefficient, $r = -0.5, \ 0, \ 0.5$.  

As anticipated, the large mass scatter in these proxies means 
that predictions for the mean and variance in  $\Ngal$ shift considerably as $r$ is varied.  
The modal value of $\Ngal$ is $62$ for $r=-0.5$, increasing to $86$ at $r=0.5$, while the scatter in
$\ln \Ngal$ drops from $0.60$ to $0.38$ for these two cases.   These effects conspire to 
dramatically change the $2.5 \sigma$ lower limits for $\Ngal$, with values of $14$ 
and $33$ for $r = -0.5$ and $0.5$, respectively.  

Observational data demonstrate a large scatter between the richness, $\Ngal$, and 
X-ray luminosity \citep[\eg][]{Rykoff0709.1158, Andreon1001.4639}, but a new approach to estimating 
optical richness from multi-color photometry offers significant improvement.  The $\lambda$ richness 
estimator \citep{Rykoff1104.2089} uses a probabilistic membership trained on spectroscopic 
calibration of the red sequence in multi-color space.  This richness estimator 
shows considerably smaller scatter in 
$\Lx$ compared to $\Ngal$ and the technique has been extended for use as a 
cluster finder in photometric surveys \citep{Rykoff1303.3562}.  Based on matching clusters found in the 
SDSS DR-8 sample to known X-ray clusters, \citet{Rozo1303.3373} demonstrate that the implied mass 
scatter of the $\lambda$ richness measure is $\sim \! 25 \%$, considerably reduced relative to 
the original $\Ngal$ richness.  

The lower panel of of Figure~\ref{fig:NgalGivenLxYsz} demonstrates how the use of low 
scatter mass proxies can significantly tighten the conditional likelihood of optical richness.  
Shown is the likelihood, $p(\lambda | \Ysz)$ at $z=0.23$ for $D_A^2 \Ysz = 3 \times 10^{-5} {\rm Mpc}^2$,
a value  that selects roughly the same mass as the $\Lx$ choice used in the top panel.  We 
assume $23\%$ mass scatter in $\lambda$ and $8.8 \%$ mass scatter in $\Ysz$.  Relative to the
top panel, the scatter in $P(\lambda | \Ysz)$ is reduced by more than a factor of two, and  the 
sensitivity of $\langle \lambda | \Ysz  \rangle$ to the correlation coefficient is also weakened. 

% In appendix \ref{conditional_proba_fits}, we show precisely the accuracy gained on $P(\ln S_b | S_a)$ with the extended models.

%%%%%%%%%%%%%%%%%%%%%%%%%%%%%%%%%%%%%%%%%%%%%%
\section{Discussion} \label{sec:discussion} 

We have emphasized an application of the model to the massive halos 
that host galaxy clusters at late cosmic times, but the mathematical framework 
is general, so the model could be applied at earlier epochs to 
describe phenomenology associated with the high-mass end of the mass function.  
Galaxies and quasars at redshifts of a few or early star formation at redshifts of tens are potential 
applications.  The key requirements are observables or properties that of halos that scale as 
power-laws with halo mass in the mean, with variability described by a log-normal covariance.  

Applied to groups and clusters, the third-order model is essentially global in scope.   
Compared to local Tinker convolution estimates, the cubic approximation achieves better 
than $10\%$ accuracy over nearly the whole signal ranges covered by current observations, 
and for redshifts 
$z < 1.5$.  This level of accuracy is comparable to the current level of systematic uncertainty 
in the mass function derived from simulations, particularly when the effects of baryon physics 
are included \citep{Stanek0809.2805, Martizzi1307.6002, Cusworth1309.4094, Cui1402.1493}.   

The pivot location sets the range of accuracy for the lower-order 
approximations.  Since the third-order approximation is 
based on a Taylor expansion of $e^{-\betathree \mu^3/6} \simeq 1 - \betathree \mu^3/6$, 
where $\mu = \ln(M/M_p)$, the model breaks down as $\mu^3 \rightarrow 6/\betathree$.  
For the pivots chosen here, this occurs only for very rare, massive systems with 
$M_{500} \gta 10^{15} \msol$.   To achieve the widest possible dynamic range, the 
second-order expressions could be interpolated using multiple pivot points, $M_{p,k}$, requiring 
values of $\betaone(M_{p,k}, z)$ and $\betatwo(M_{p,k}, z)$ to be provided.  For light-cone 
applications, these derivatives could be modeled as low-order polynomials in redshift.  

The pivot masses we employ at the two demonstration redshifts correspond 
closely to those satisfying a fixed sky surface density condition, $dN(> \! M)/dz = {\rm const.}$, 
in a $\Lambda$CDM cosmology \citep{Evrard0110246, Mortonson1011.0004}.  For cluster survey 
applications, this would seem a natural choice.  

While power-law scaling with log-normal covariance is supported for intrinsic properties 
of clusters, the available evidence is often limited.  
In particular, covariance among 
different signals is poorly understood \citep{Maughan1212.0858} 
and there are few constraints as to whether 
the slope and variance of a particular signal's scaling with mass is indeed constant with 
mass and redshift, as is assumed here 
\citep[\eg][provide a hint of evidence for curvature in the $\Lx$--$M$ relation]{BalagueraAntolnez1207.2138}.  Redshift dependencies are easily incorporated into the 
existing framework by writing the slopes, $\boldalpha(z)$, intercepts, $\boldpi(z)$, and covariance, $C(z)$, as explicit 
functions of $z$.   Extensions to the model that incorporate weak mass dependence in the scaling 
parameters are also possible.  

The model covariance can be interpreted as that among intrinsic properties of halos, but 
signals observed from real clusters inevitably include projection effects and noise, 
including potential bias, associated with signal detection and characterization.  These effects, 
particularly projection for SZ and optical signatures, deserve further exploration 
\citep{NohCohn1204.1577, White1005.3022, Angulo1203.3216}.

%%%%%%%%%%%%%%%%%%%%%%%%%%%%%%%%%%%%%%%%%%%%%%
\section{Summary} \label{sec:summary} 

Using polynomial log-space approximations to the high-mass end of the cosmic mass 
function, we present analytic forms for statistics of multi-observable properties 
of the high-mass halos that host groups and clusters of galaxies.  The model employs 
scaling laws between observables and mass that are power-law in the mean with 
log-normal covariance.  

The model provides quick estimation of cluster counts as a function of multiple observables and 
calculation of conditional likelihoods for observable-selected samples, both of which are directly 
relevant for joint survey analysis or follow-up observations.  By comparing to a locally-convolved 
Tinker mass function, we show that the first-order model is generally accurate within a narrow 
range near the pivot mass, except for very high mass-scatter proxies.  The 
second and third-order extensions provide increasingly wider coverage 
in observables irrespective of the mass scatter.  The third-order model is nearly global in scope.  

The mass variance in a particular observable determines many expected features, as does 
the covariance between pairs of observables at fixed mass.  As multi-wavelength surveys and dedicated 
follow-up campaigns provide increasingly rich, uniform samples of clusters, opportunities to apply 
this model to better constrain the statistical properties of massive halos will become apparent.  
Such knowledge will provide useful constraints on the physical processes that govern baryon 
evolution in massive halos.

\section*{Acknowledgments}   We thank the organizers of the ``Monsters Inc.'' workshop on clusters 
at KITP, supported in part by the National Science Foundation under Grant No. PHY05-51164, 
where this work initiated.   We thank Eduardo Rozo and Jim Bartlett for useful suggestions.  
The mass function calculations were obtained from http://hmf.icrar.org/ using source code 
developed by Steven Murray and collaborators at http://github.com/steven-murray/hmf/.  
AEE acknowledges NASA NNX07AN58G and the Mairie de Paris 
``Research in Paris'' program for support, and thanks Gary Mamon and Joe Silk for 
sabbatical hosting assistance at IAP.  PA thanks A. Milliken and the 
Michigan Center for Theoretical Physics for hospitality and acknowledges support from the 
French Fonds de Solidarit\'e et de D\'eveloppement des Initiatives \'Etudiantes (FSDIE).   

\appendix

\section{Third-order model}

\subsection{Multi-property halo number density}

The accuracy gained in going from $\none(s)$ to $\ntwo(s)$ motivates a third-order approach.   
We thus now consider
\begin{equation}
\nthree(\mu) = A e^{-\betaone \mu - \frac{1}{2}\betatwo \mu^2 - \frac{1}{6} \betathree 
\mu^3} \ ,
\end{equation}

\noindent
where $\betathree = - [(d^3/d\mu^3) \ln n_{\rm exact}](\mu = 0) > 0$.  We could not find 
a closed form  solution at this order, so we instead consider the approximation that limits 
the mass range to be near enough to the pivot point so that 
\begin{equation} \label{approx}
\nthree(\mu) \simeq \ntwo(\mu) \left(1 - \frac{1}{6} \betathree \mu^3 \right) .
\end{equation}
\noindent
% which is valid for $\frac{1}{6} \betathree \mu^3 \ll 1$.\\

Convolving this approximated form of $\nthree(\mu)$ with $P(\mathbf{s}|\mu)$ 
yields
\begin{equation} 
\label{eq:n3}
\nthree(\mathbf{s}) = \ntwo(\mathbf{s})\left\lbrace 1 - \frac{\betathree}{2} \left
( \sigma_{\mu|\mathbf{s},2}^2  \langle \mu | 
\mathbf{s} \rangle_2   + \frac{1}{3} \langle \mu | \mathbf{s} \rangle_2
^3 \right) \right\rbrace \ .
\end{equation}
One can see clearly that the signal range will be limited from above by the requirement 
that the space density be non-negative.

\subsection{Selecting on an observable property}

We have $P_3(s_b|s_a) = \nthree(s_a,s_b)/\nthree(s_a)$, 
which, using the previous formula, gives 
\begin{eqnarray}
\lefteqn{P_3(s_b|s_a) \ = \ P_2(s_b|s_a)} \\
& & \times \ \frac{1 - \frac{\betathree}{2} \left
( \sigma_{(\mu|a,b),2}^2  \langle \mu | s_a,s_b 
\rangle_2  + \frac{1}{3} \langle \mu | s_a,s_b \rangle_2^3 \right)}{1 - 
\frac{\betathree}{2} \left( \sigma_{\mu|a,2}^2  
\langle \mu |s_a \rangle_2   + \frac{1}{3} \langle \mu |s_a \rangle_2^3 
\right)} . \nonumber
\end{eqnarray}

We can calculate analytically the mean and variance of $P_3(s_b|s_a)$, 
wich gives
\begin{eqnarray}
\label{mean_P3(s2 | s1)} 
\lefteqn{\langle s_b | s_a \rangle_3  \ = \ K_1 J_1 - K_2 C(J_2 + D J_1)} \\ 
& & - \ K_3 C^3 (J_4 + 3 D J_3 + 3 D^2 J_2 + D^3 J_1) \nonumber \\
\nonumber \\
\label{sigma_P3(s2 | s1)} 
\lefteqn{\sigma_{b|a,3}^2  \ = \ K_1 J_2 - K_2 C(J_3 + D J_2)}  \\
& & - \ K_3 C^3 (J_5 + 3 D J_4 + 3 D^2 J_3 +  D^3 J_2)  - \langle s_b | s_a \rangle_3^2 , \nonumber
\end{eqnarray}
 
\noindent
where
\begin{flalign}[left=\empheqlbrace]
&J_n  \ = \ \int_{-\infty}^{+\infty} s_b^n P_2(s_b|s_a) ds_b \\
& \ \ \ \ = \ \frac{1}{\sqrt{\pi}} 
\sum_{k=0}^{\lfloor n/2 \rfloor} \binom{n}{2k} M^{n-2k} 
(2\sigma^2)^k \Gamma(k+1/2) \\
&M  \ = \ \langle s_b | s_a \rangle_2 \\
& \sigma  \ =  \ \sigma_{b|a,2} & &
\end{flalign}
\noindent
( In particular : $J_0 = 1 \ , \ \ \ J_1 = M \ , \ \ \ J_2 = \sigma^2 + M^2$  ) \\
\noindent
and
\begin{flalign}[left=\empheqlbrace]
&C \ = \ \frac{X \ \sigma_{(\mu|a,b),1}}{1 - \rab^2} \\
&D \ = \ \frac{1}{X} [ Y - X\pi_b - \betaone(1-\rab^2) ] \\
&X \ = \ \frac{\alpha_b}{\sigma_b^2} - \frac{\rab \alpha_a}{\sigma_a\sigma_b} 
\\
&Y \ = \ (s_a - \pi_a) \left( \frac{\alpha_a}{\sigma_a^2} - \frac{\rab \alpha_b}{\sigma_a \sigma_b} \right) & &
\end{flalign}
\begin{flalign}[left=\empheqlbrace]
&K_1  \ = \ \left[ 1 - \frac{\betathree}{2} \left( \sigma_{\mu|\mathbf{s},2}^2  \langle 
\mu | \mathbf{s} \rangle_2   + \frac{1}{3} 
\langle \mu | \mathbf{s} \rangle_2^3 \right) \right]^{-1} \\
&K_2 \ = \ K_1 \frac{\betathree}{2} \ \sigma_{(\mu|a,b),1} \ x_{a,b}^2 \\
&K_3 \ = \ K_1 \frac{\betathree}{6} \ x_{a,b}^3 \ . & &
\end{flalign}

\bibliographystyle{mn2e}

\bibliography{clusters}

\end{document}